\documentclass[12pt, conference, compsocconf, onecolumn]{IEEEtran}

\usepackage{hyperref}
\usepackage{xcolor}

\usepackage{listings}
\usepackage{xspace} 
\usepackage{graphicx} 
\usepackage{algorithm2e} 
\SetAlFnt{\footnotesize}

\usepackage[tight,footnotesize,hang,center]{subfigure}

\graphicspath{{./figures/}}
\DeclareGraphicsExtensions{.pdf}

\newcommand{\floe}{$\mathcal{F}{\ell}{o}{\varepsilon}$\xspace}

\begin{document}

\title{{\emph{Floe}:} A Continuous Dataflow Framework for Dynamic Cloud Applications}
\author{\IEEEauthorblockN{Yogesh Simmhan and Alok Kumbhare}
\IEEEauthorblockA{University of Southern California\\
Los Angeles CA 90089\\
\{simmhan, kumbhare\}@usc.edu}
}

\maketitle

\begin{abstract}
Applications in cyber-physical systems are increasingly coupled with online instruments to perform
long-running, continuous data processing. Such ``always on'' dataflow applications are dynamic,
where they need to change the application’s logic and performance at runtime, in response to
external operational needs. \floe is a continuous dataflow framework that is designed to be adaptive
for dynamic applications on Cloud infrastructure. It offers advanced dataflow patterns like BSP and
MapReduce for flexible and holistic composition of streams and files, and supports dynamic
recomposition at runtime with minimal impact on the execution. Adaptive resource allocation
strategies allow our framework to effectively use elastic Cloud resources to meet varying data
rates. We illustrate the design patterns of \floe by running an integration pipeline and a tweet
clustering application from the Smart Power Grids domain on a private Eucalyptus Cloud. The
responsiveness of our resource adaptation is validated through simulations for periodic, bursty and
random workloads.\end{abstract}


\section{Introduction}

Scientific and engineering applications have grown more data intensive and distributed over the past
decade. Compositional models such as workflows and stream processing systems have proliferated to
help design and execute complex application pipelines easily. As instruments and sensors become more
pervasive -- approaching the ubiquitous ``Internet of Things'' -- these dataflow applications tend
to operate over continuous data, run for extended periods, and evolve as the domain needs change.
Recently, Clouds have emerged as a popular 
cyber-infrastructure to orchestrate and execute such dataflows. Their elastic scaling
offers a unique runtime for dynamic dataflow applications. In the class of dynamic, distributed and
data-intensive applications~\cite{d3science}, the dynamism property is less well
studied. Dynamism manifests both in the continuous and varying nature of \emph{data}, and
in the evolving \emph{application} logic and performance.

Consider the Smart Power Grid domain, a cyber-physical system where dataflow applications process
real-time data from thousands of devices for grid
operations~\cite{Simmhan:buildsys:2011}. \emph{Information integration} from diverse sources and
\emph{analytics} on energy demand are two canonical applications here. Firstly, the integration
pipeline constantly acquires data from smart meters, building sensors, and weather services,
that arrive as events, XML documents and archived files, and at
frequencies that vary from 1/minute to 1/day (Fig.~\ref{fig:sampleapps}(a)). As sensors
expand and sampling frequencies are tuned, the performance profiles of the dataflow changes,
requiring runtime adaptation to meet latency goals. The dataflow itself will also change for bug
fixes and new cleaning techniques; but, the always-on nature of power grids requires in-place
updates to a running dataflow's logic. This exemplifies online \emph{application dynamism}, both in
the composition and in performance. 

Secondly, analytics help train demand forecasting models like regression tree over historical energy
data using variations of the MapReduce dataflow~\cite{Yin:mapreduce:2012}. Complex event
processing~\cite{Zhou:ef:2010} offers online analytics over real-time streams to detect situations like
a mismatch between forecast power demand and expected supply.  Graph analytics over 
social networks help target customers for real-time energy
reduction~\cite{petkov2011motivating}. Bulk Synchronous Parallel (BSP) dataflows~\cite{malewicz2010pregel} are gaining
traction for such graph algorithms. These \emph{diverse and specialized dataflow
abstractions} coexist within one domain and also need to coordinate with each other
(e.g. integration $\rightarrow$ forecasting $\rightarrow$ detection $\rightarrow$ correction).



These expose two key gaps in existing dataflow frameworks: (1) Missing support for application
dynamism, and (2) Limited capabilities for holistic dataflow composition.  Contemporary scientific
workflow and stream processing systems~\cite{Deelman2002,Ludscher2005,neumeyer2010s4,amini2006spc} allow eScience applications to
be composed as directed task graphs, and launched for execution on computational resources such as
workstations, HPC and Cloud resources. However, these frameworks do not support changes to the
composition once the dataflow has been launched. Changes may include updates to the task logic or
the data flow routing. Rather, the dataflow has to be stopped, recomposed, and (re)launched, or
alternatively, an updated duplicate dataflow launched while the older one is quiesced. These can
lead to lost data from unbuffered real-time streams, and data inconsistencies in the absence of
tight coordination. There is better support for adapting to data dynamism~\cite{neumeyer2010s4} than
application dynamism. But effective utilizing of Cloud elasticity to meet runtime performance
metrics for continuous dataflows is still nascent.

Directed cyclic or acyclic dataflow task graphs, offered by workflow and stream processing
frameworks, allow a large class of applications to be composed. In practice, advanced dataflows
constructs such as BSP and MapReduce are not natively supported.  While MapReduce is a directed
acyclic graph (DAG), the critical ``shuffle'' step, where mapper outputs are grouped by key and
routed to specific reducers, is not efficiently handled by generic workflow frameworks, if at all.
Like Hadoop for MapReduce~\footnote{Apache Hadoop, \url{hadoop.apache.org}}, BSP too requires
special frameworks like Apache Hama~\footnote{Apache Hama, \url{hama.apache.org}}.  Efforts to
incrementally extended these frameworks with compositional semantics, such as iterative MapReduce
(e.g. Twister~\cite{twister}), still fail to offer rich dataflow abstractions. Else, users are
forced to adopt and maintain multiple frameworks.  Our own earlier work~\cite{Zinn:ccgrid:2011}
addressed some issues related to flexible dataflows by integrating streams, collections and files as
first-class data models (i.e. types of edges) that could be used in dataflows. 
This paper investigates more advanced patterns like BSP, but goes further to 
address issues of runtime dynamism in the application logic and adaptive resource allocations on
Clouds to meet application performance needs.

In summary, we make three contributions in this paper: 
\begin{enumerate}
\item We propose \emph{design paradigms} for a continuous dataflow framework to support
dynamic, long-running applications on Clouds. Besides basic dataflow patterns, the design
is extensible to streaming MapReduce and BSP. Our ability to change the task logic at runtime
facilitates application dynamism. (Sec.~\ref{sec:design})
\item We present the \emph{architecture} of \floe, a continuous dataflow framework for
  Cloud infrastructure, that supports the design paradigms for dynamic applications and uses
  \emph{adaptive resource allocation} to meet dataflow latency goals under dynamic conditions. (Sec.~\ref{sec:arch})
\item We validate our design and empirically evaluate our framework implementation on a Eucalyptus private
  Cloud for 
  two real-world Smart Grid applications. In addition, our adaptive allocations are validated
  through simulations. (Sec.~\ref{sec:eval})
\end{enumerate}

\section{Design Paradigms of \floe}
\label{sec:design}

\subsection{Flexible Dataflow Abstractions}
Workflow and dataflow patterns have been well examined, but scalable workflow systems, in practice,
support a compact set of patterns. Here, we discuss basic and advanced patterns that go toward
holistic dataflow compositional.

\textbf{Continuous Dataflow Model.} 
A continuous dataflow design is foundational to \floe, where applications are composed
as a directed graph with tasks -- known as \emph{pellets} -- as vertices and data channels
connecting them. Pellets are user's application logic that implement one of several
\texttt{compute()} interfaces that are provided by the framework. Pellets expose one or
more named input and output ports that can receive or emit a stream of messages. These messages may
be small, serialized Java objects or large files.

\begin{figure}
\includegraphics[width=\columnwidth]{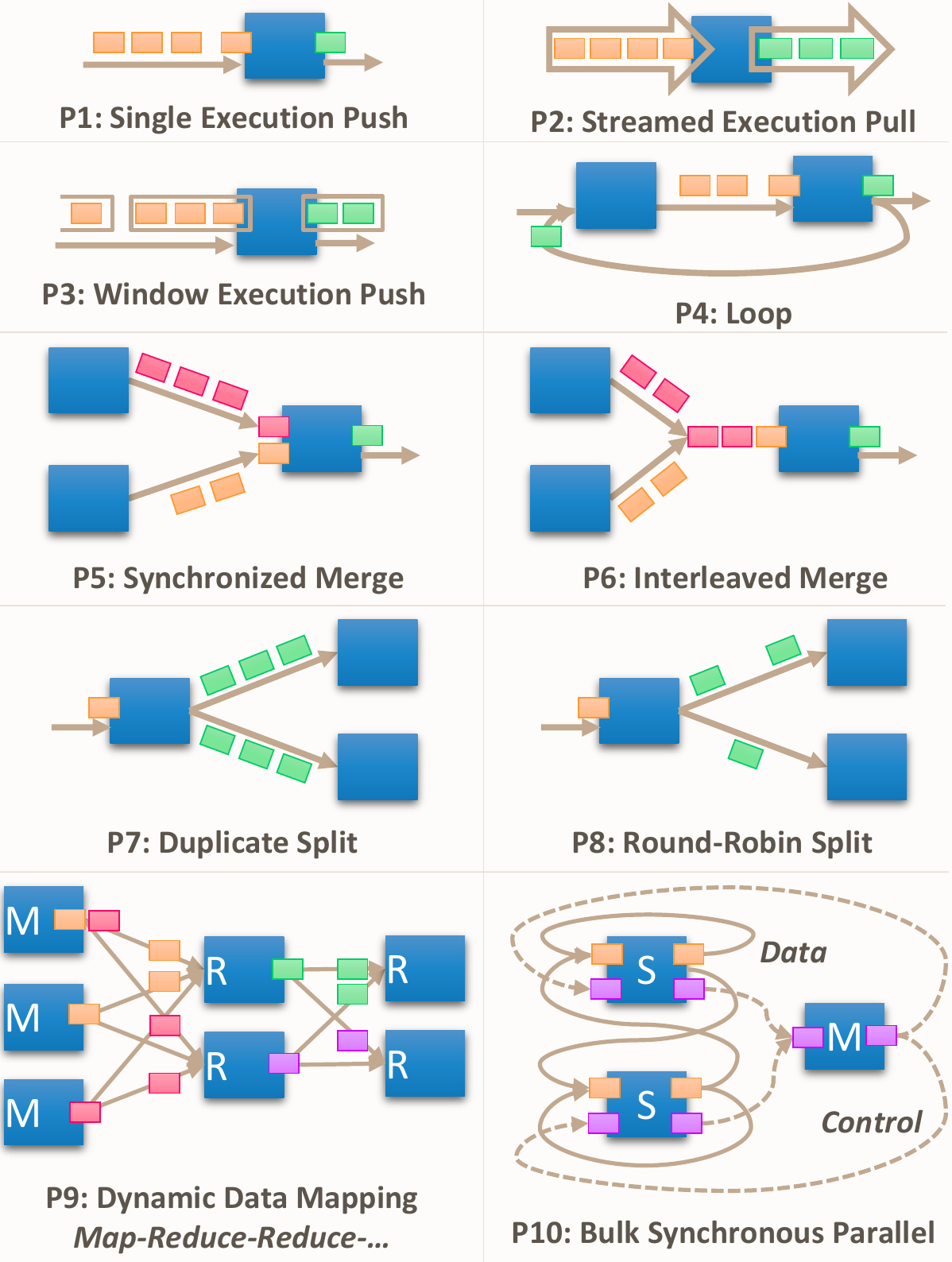}
\caption{Subset of Dataflow Patterns supported in \floe}
\vspace{-0.25in}
\label{fig:pattern}
\end{figure}

\floe offers both a push and a pull model of \emph{triggering} pellet execution on incoming
messages. In a \emph{push} interface, the framework invokes the pellet's \texttt{compute()} method for each
available message on the input port (Fig.\ref{fig:pattern}, P1). Pellet using a \emph{pull} interface are
designed for stream execution and their \texttt{compute()} method can access an \texttt{iterator} of
messages (Fig.\ref{fig:pattern}, P2). Output messages can likewise be \emph{return}ed by the
\texttt{compute()} interface or written to an output \texttt{emitter} stream. While every input
message should generates one output message in a push model, pull pellets may consume zero or more
messages to emit zero or more messages. For pellets with multiple ports, the inputs and outputs are
a \texttt{map} (tuple) object whose keys are the port names and values are the corresponding message
on that port (Fig.\ref{fig:pattern}, P5). Pellets can also receive a collection of messages that
fall in a time or count window, the width of which is specified at graph composition time by
the user (Fig.\ref{fig:pattern}, P3).

Push pellets are \emph{implicitly stateless} while those that pull may retain local state. \floe
provides pellets the ability to explicitly store and retrieve a \emph{state object} that can be
retained across pellet invocations. Using an explicit state object allows the framework to (in
future) offer resilience through transparent checkpointing of the state object and resuming from the
last saved state and the input messages available then.  Support for large messages and a push mode
of execution allows our continuous dataflow design to also be used for traditional batch processing
workflows that operate on files.

\textbf{Basic Dataflow Abstractions.} 
Pellets can be composed into directed graphs where ports are wired to each other to indicate dataflow
between them. \emph{Control flow} constructs such as \texttt{if-then-else} or \texttt{switch} can be
easily implemented through pellets with multiple output ports, where the user logic emits an output
message on only one of them. \floe graphs can also have cycles, so that an output port can be
connected to the input port of a ``preceding'' pellet. This allows iteration constructs like
\texttt{for-loops} to be designed (Fig.\ref{fig:pattern}, P4).

\floe graphs support different patterns for aggregating messages from multiple pellets. A
\emph{synchronous merge} (Fig.\ref{fig:pattern}, P5) aligns messages that arrive on different input
ports to create a single message tuple map, indexed by port name, that can be pulled by or pushed to
the pellet. Alternatively, pellets multiple input edges can be wired to a single port as an
\emph{interleaved merge} (Fig.\ref{fig:pattern}, P6) where messages from either input edges are
accessible on that input port immediately upon their arrival.

\emph{Task parallelism} is achieved explicitly by wiring output ports of a pellet to different
pellets. When an output port of a (source) pellet is split to multiple (sink) pellets, users can
decide at graph composition time an output message should be \emph{duplicated} to all outgoing edges
(Fig.\ref{fig:pattern}, P7) or sent to only one. In the latter case, a default \emph{round-robin}
strategy is used to select the edge as load balancing (Fig.\ref{fig:pattern}, P8) but a more
sophisticated strategy can be employed in future (e.g. depending on the numbers of messages pending
in the input queue for the sink pellet).

Every pellet is \emph{inherently data parallel}. The \floe framework transparently creates multiple
instances of a pellet to operate on messages available on a logical input port. The number of
instances created is determined by our optimizations, discussed later. Pellet instances also share
the same logical output port. One side-effect of this is that the output messages may not be in the
same sequence as the input, since the pellet instances may complete out of order. Users can
explicitly set an option that forces a pellet to operate sequentially without data parallelism to
ensure messages flow in order.





\textbf{Advanced Dataflow Abstractions.} 
The basic dataflow patterns can be composed to form complex applications and advanced
constructs. For e.g., \emph{Bulk Synchronous Parallel (BSP)} model has seen a revival off-late for
large scale graph algorithms on Clouds~\cite{malewicz2010pregel}. BSP is an 's' stage (also called superstep)
bi-partite dataflow where each superstep has 'm' identical pellets that operate concurrently and
emit messages to other pellets in that superstep. The message transfer itself is synchronous at
superstep boundaries, and made available as input to pellets once the next superstep starts. The
number of supersteps is decided at runtime. We can compose a native BSP model using the basic \floe
patterns using a 's' fully pellets with all their output ports connected to each others' input ports
(Fig.\ref{fig:pattern}, P10). In addition, a manager pellet acts as a synchronization point to
determine when a superstep is completed by all pellets. So ``data'' messages on the input port of
superstep pellets are gated by a ``control'' message from the manager pellet on another the input port.

\emph{MapReduce} is another advanced dataflow pattern, constructed as a two-stage bi-partite graph from 'm'
mapper tasks to 'r' reducer tasks. Since its ``shuffle'' between Mappers
and Reducers does not naturally fit the basic patterns -- a shortcoming of other dataflow
frameworks-- we introduce the notion of \emph{dynamic port mapping} during a split pattern. This
allows continuous MapReduce+ to be composed, with one Map stage and one 
or more Reduce stages (Fig.\ref{fig:pattern}, P9).  Map and Reduce pellets are wired as a bipartite
graph similar to a MapReduce dataflow, and they both emit messages as $<$key,value$>$
pairs. However, rather than use duplicate or round-robin split of messages from the output port of a
Map pellet to the input port of the Reducer pellets, the \floe framework performs a hash on the key
to dynamically select one of the edges on which message passes, similar to Hadoop. The hash ensures
that messages from any Map pellet having the same key reaches the same Reduce pellet.

The Map and Reduce pellets can be used in any dataflow composition. This approach generalizes the
pattern beyond just MapReduce, and allows even iterative MapReduce composition (but does away with
the, often, unnecessary Map stage for the second and subsequent iterations). Also, our streaming
version allows Reducers to start before all Mappers complete, and also allows operation over
incremental datasets as they arrive rather than in batch mode.  Pellets can emit user-defined
``landmark'' messages to indicate when a logical window of message streams have been processed to
allow the reducer pellets to emit their result, say, when they are doing aggregation.

\subsection{Application Dynamism}
Continuous dataflow applications by definition run constantly, processing data streams. As a result,
making updates to the application, say to fix a bug or the change the application's behavior, can
disrupt its execution; this may not be acceptable to operational dataflows. In \floe, we explore two
types of application dynamism, and mechanisms to mitigate disruption and ensure consistent results.

\textbf{Dynamic Task Update.}
A common form of application dynamism is when the logic for a single pellet needs to change during a
during a dataflow execution. \emph{Stopping} and \emph{starting} the entire dataflow, while commonly
supported, can lead to incorrect results if state is not preserved, and lead to time consuming resource
reallocation. Even if we can \emph{pause} a dataflow retaining the state of each pellet, it can
cause data losses unless every data stream in-flight is be buffered -- this can be resource
intensive for high-frequency streams. 

We take a more nuanced approach to dynamic task updates in \floe and allow individual pellets to be
updated in-place without halting the execution of other pellets. There are several issues to
consider here. An in-place task update requires that the number of ports in the old and new pellets
has to be the same, as does their interfaces. Otherwise, this degenerates to dynamic dataflow
update, discussed next. Internal state held by the pellet is retained if it is a stateful pellet --
its state object will survive the update. Messages pending in its input ports will be retained and
available to the new pellet. Depending on the degree of consistency required, task updates can be
synchronous or asynchronous. If \emph{synchronous}, messages being processed will finished to
completion and messages already generated in its output port will be delivered before the new pellet
is instantiated. The new pellet can optionally send an ``update landmark'' message to downstream
pellets in the \floe graph to notify them of a new logic in place. In an \emph{asynchronous} update,
messages being processed by pellet instances will continue to completion even while instances of the
updated pellet are activated and get access to the input messages. Output messages from both the old
and the new pellet instances may be interleaved.

The advantage of this model is the zero downtime in case of asynchronous task updates, and minimal
downtime for synchronous ones, limited to the time needed to finish processing input messages that
have already been retrieved by a pellet instance. For pellet instances that take a long time to
process a single message, the \floe framework can optionally deliver an \texttt{InterruptException}
to the pellet user logic to allow it to save its state and conclude processing its message.

\textbf{Dynamic Dataflow Update.}
Changes to the structure of a dataflow at runtime can have a more significant impact. One common
scenario is when a sub-graph of the dataflow is added, removed or replaced. These can be
reduced to the previous case of task update, where instead of one task, multiple individual pellets
are added, removed or updated in-place from the dataflow \emph{in a coordinated manner}. Here,
synchronization goes beyond just a pellet and requires all pellets in the sub-graph to be added,
removed or updated simultaneously. Here, the synchronization cost can be higher than before
since the slowest pellet update becomes the bottleneck. While \floe supports these ``structural'' updates, the
user still needs to consider the impact on the application's consistency in such a scenario and use the
``update landmark'' (or other custom mechanisms) to notify downstream pellets of the change. 

A more sophisticated mechanism for future, that better addresses consistency issues, is to perform a
cascading update to the dataflow graph when a sub-graph is being replaced. Here, an
``update tracer'' message would traverse from the source pellet(s) of the sub-graph to the sink
pellet(s), and perform an in-place update of each pellet when it reaches it. This form of ``wave''
propagation ensures that a clear distinction can be made between data streams emitted before an
update and those emitted after. Pellets will still need to consider the impact of the update on
stateful pellets.  

\section{Adaptive Architecture}
\label{sec:arch}

\begin{figure}
\centering
\includegraphics[width=\columnwidth]{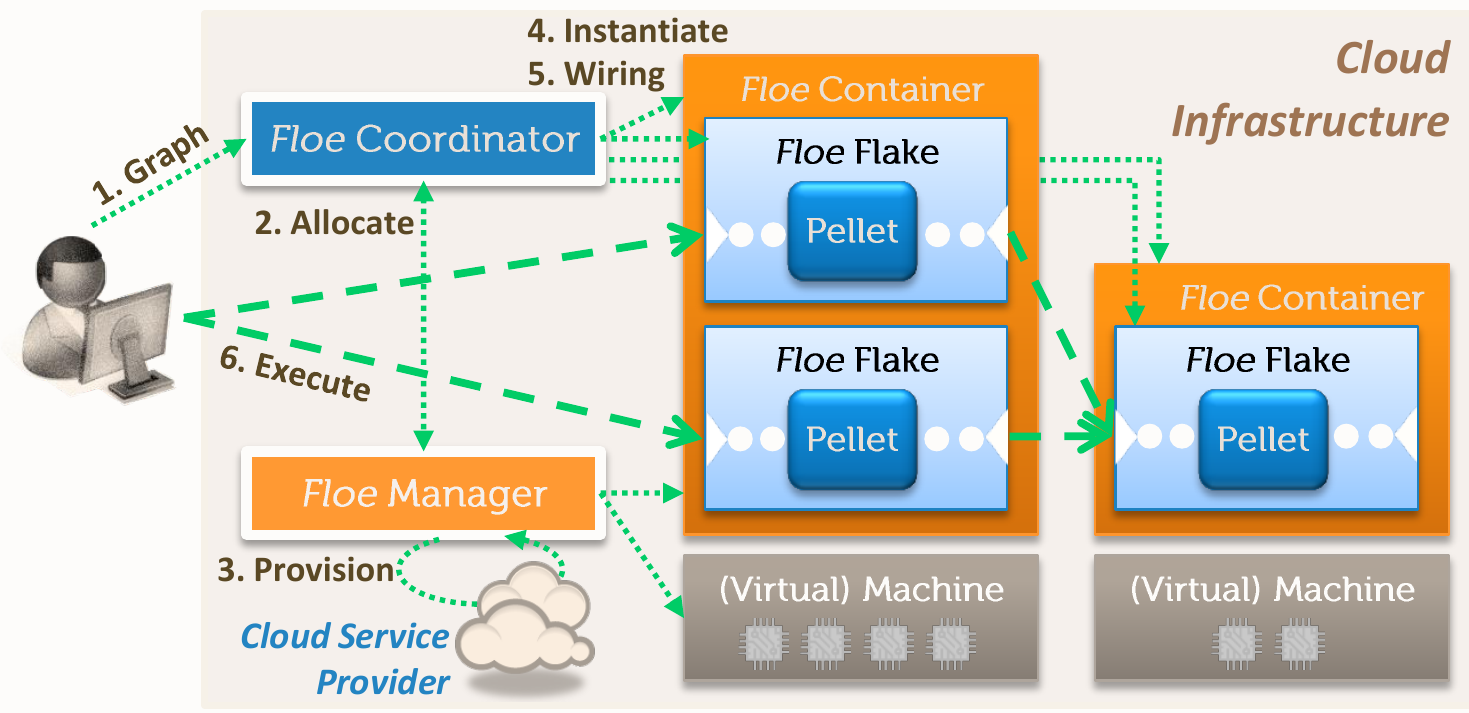}
\caption{The \floe Architecture}
\vspace{-0.25in}
\label{fig:arch}
\end{figure}


The \floe architecture incorporates the design patterns and dynamism we introduced, and is
implemented on Cloud infrastructure. Pellets are Java classes that encapsulate the user's application
logic and implement one of several interfaces that determine the ports and the triggering mechanism.
Input and output messages are Java serializable objects and may include file streams.  \floe
applications are composed as a directed graph, described in XML, where vertices are pellets
identified by their qualified class name and dataflow edges identify the input and output ports of
the source and sink pellets they connect. Edges and ports can be annotated with design pattern
properties like data parallel execution, statefulness, window width and duplicate/round-robin
split. Edges can also state if messages are transmitted synchronously from a source to a sink
pellet, or is asynchronously retrieved by the sink from the source. These allow flexibility during
application composition rather than deciding at pellet development time.



The \floe Cloud execution framework contains an application runtime and a resource runtime. The
\emph{Coordinator} and \emph{Flake} are responsible for the application runtime at the \floe graph
and pellet granularities, respectively, while the \emph{Manager} and \emph{Container} handle
resource runtime at the Cloud datacenter and Virtual Machine (VM) (Fig.~\ref{fig:arch}).  

\floe adopts a component model of application execution \footnote{Common Component Architecture,
\url{http://www.cca-forum.org}} so that a single centralized dataflow orchestrator is not a
bottleneck. A \textbf{flake} is responsible for executing a single pellet in the \floe graph and
coordinating dataflow with neighboring flakes.
A flake has an input and an output queue for buffering de/serialized messages to/from its pellet. Depending
on the design pattern annotation, the flake aligns input messages from multiple ports or windows
messages and makes it available to the pellet. Flakes also create data parallel pellet instances for
consuming messages, either once for pull pellets that iterate over input messages or once per
message/tuple for push pellets. Output messages from a pellet instance are placed in the output
queue and the flake transmits messages to the sink pellets (flakes) directly, following the split
rules, or supports asynchronous retrieval by a sink pellet (flake). \floe offers multiple transport
channels, including direct socket connections between flakes and (in future) Cloud queues, depending
on performance and resilience requirements.

Each flake runs on one Cloud VM, and resource provisioning within a VM is done by the
\textbf{container}. Container manage instantiation of one or more flakes in a VM and allocates
specific numbers of CPU cores to them using Java 7's \emph{ForkJoinPool} mechanism. Pellet instances
created by a flake are restricted to run on the allocated core(s), allowing light-weight
multi-tenancy within a VM. The ratio of the number of pellet instances to the number of cores is set
to a static value ($\alpha=4$, presently). However, the numbers of cores allocated to a flake can be changed
dynamically during execution time using control interfaces, giving fine-grained resource control.

The \textbf{coordinator} parses the \floe graph provided by the user and acquires cores on different
VM containers to instantiate flakes corresponding to pellets in the graph. The coordinator
negotiates with the \textbf{resource manager} for core allocation and the container location. The
manager interacts with the Cloud service provider to acquire and release VMs on-demand based on the
current need. Currently,
we support the Eucalyptus Cloud fabric that is also compatible with Amazon AWS. Once flakes are
active, the coordinator ``wires'' and activates the flakes based on the dataflow graph and design
pattern annotations. The wiring is done as a bottom-up breadth-first search traversal of the
dataflow (ignoring loops) to ensure that upstream pellets are not active and generating messages
before downstream pellets are wired and active. The coordinator returns the input port endpoint of
the initial flake(s) to the user to allow her to pass initial inputs to the dataflow, if required.  Task and
dataflow dynamism is handled by the coordinator, and it interacts with the flakes to pause, swap and
resume the pellets. The coordinator, manager, container and flake expose REST web service endpoints
for these management interactions.

\textbf{Resource Adaptation Strategies.}
The \floe graph can be statically annotated with the number of CPU cores allocation for
each pellet. The coordinator will use this information to request existing or newly
instantiated containers from the manager using a best-fit algorithm. Multiple flakes may be packed
in the same container, and these flakes can span multiple \floe graphs for multi-tenancy.
However, given the data dynamism, where stream rates may change over time, static core allocation
can lead to over or under provisioning by the user.

We investigate three resource adaptation strategies, static, dynamic, and hybrid, to handle
different stream rates and periodicity, and processing latencies for pellets in the
dataflow. Performance metrics we 
optimize for are to ensure that pellets can \emph{sustain} continuous dataflow processing at the
input data rate, and to bound the end-to-end processing \emph{latency} for specific processing time
windows. These strategies use instrumentation present within flakes for monitoring their queue
lengths and average message latencies, and are used to adapt the core allocation to a flake at runtime.

The \textbf{Static look-ahead} strategy assumes that the user (an ``oracle'') has picked the
best static allocation of cores for each pellet in the dataflow based on historical observations of
its workload and data rates. Static allocation can be well suited when the workload is deterministic
and periodic where a sample benchmark run can be extrapolated. For each of $n$ pellets along the
dataflow's critical path, if $l_{1..n}$ are their per message processing latencies using one pellet
instance, $s_{1..n}$ are the selectivity ratio of the number of output messages emitted for each
input message, and $m_1$ is the number of messages that arrive at the first pellet's input port
within a period $t$, then we can achieve a sustainable rate of processing along the critical path
within a time tolerance of $\epsilon$ by statically setting the number of instances of each pellet
to $P_i$, where: $P_i \simeq (l_i \times m_i) / (t + \epsilon)$ and $m_i = m_{i-1} \times s_i$. The
corresponding number of cores, using a static instance to core ratio of $\alpha=4$, is $C_i=\lceil
P_i/4 \rceil$.

The \textbf{dynamic} strategy uses continuous monitoring of the pellet to adapt the number
of its instances. It is better suited for workloads where the data rates are slow changing but not
predictable \emph{a priori}. The goal of this strategy is to ensure that the instantaneous rate at
which input messages are arriving at a pellet can be matched by the rate at which output messages are being
emitted by that pellet's instances, after accounting for the selectivity ratio. 
The dynamic algorithm is triggered at regular intervals at which time it evaluates the rate of incoming data as well as the processing rate based on the task latency and the number of currently running pellet instances. If the data rate is observed to be greater than the data processing latency by a threshold, it increases the number of cores assigned to the pellet (and in turn the number of parallel instances by a factor of $\alpha$). On the other hand, if the data rate is less than the data processing latency by a threshold, it further checks if decreasing the number of allocated cores would violate this condition, if not, it decreases the number of allocated cores. The second check is necessary to ensure that the number of allocated cores do not fluctuate too often.

The static look-ahead strategy is well suited for periodic dataflows while the dynamic adaptation
gradually evolves with changing rates. However, dataflows rarely fall under either extremes. We
introduce a \textbf{hybrid} 
strategy that takes hints on data rate and periodicity, like the static strategy, but does not
assume that the user is an ``oracle''. Rather, it adapts and switches to a dynamic strategy when the
expected data rates start to veer off beyond a threshold.
 It switches back to the static strategy when the data rate stabilizes close to the hinted average value and the number of pending messages in the queue is below a threshold.

\begin{algorithm}
\SetAlgoLined
\KwData {DataRate D, ProcessingLatency L, \# pellet instances N}
$ \delta =  D - \frac{N}{L}$\;
\eIf {$\delta \geq \tau_1$} {
		increment allocated cores by 1\;
} {
   \If{$\delta \leq \tau_2$ and $\delta < -\frac{N-1}{L}$} {
		decrement allocated cores by 1\;
   }
}
\label{algo:dynamic}
\caption{Dynamic Adaptation of Cores for Flake}
\end{algorithm}
\vspace{-0.2in}

Dynamic adaptation requires continuous monitoring, and its effectiveness depends on the sampling
frequency (with its associated overheads). Further, the dynamic approach can only increase the core
allocation for a flake within a single VM (cross-VM elasticity and migration of flakes is planned
for future) while the static look-ahead can help split resource intensive pellets across multiple
flakes at design time. However, as our evaluation in Section~\ref{sec:eval} shows that dynamic adaptation
is more robust to dataflows with variability. The hybrid adaptation attempts to get the best of both
worlds. Our current \floe implementation supports the static strategy if provided by the user and
defaults to the dynamic strategy; implementing hybrid is part of future work.

\section{Case Study Applications and Evaluation}
\label{sec:eval}

\begin{figure*}
\centering
\includegraphics[width=0.9\textwidth]{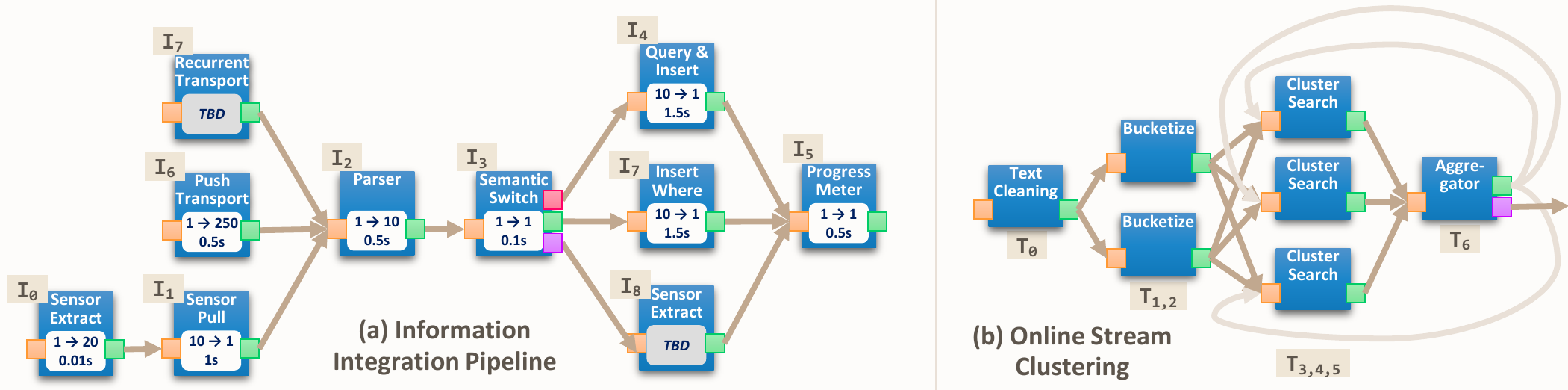}
\caption{Smart Power Grid applications using \floe. Pellets in the pipeline show
  selectivity ratio from input $\rightarrow$ output, and message processing time.}
\label{fig:sampleapps}
\end{figure*}

We present an illustrative evaluating of \floe's patterns in two representative real-world
applications from Smart Power Grids, along empirical observations of their execution on a Eucalyptus
private Cloud. Further, we also validate our resource adaptation stategies through simulation, for
three different data stream workloads.


\subsection{Smart Grid Information Integration Pipeline}


A Smart Grid information pipeline integrates diverse data from the cyber-physical system into a
knowledge database for analysis and grid operations. The pipeline used at the USC Campus Microgrid
(Fig.~\ref{fig:sampleapps}(a)), a testbed for the Los Angeles Smart Grid project, is
composed of pellets that periodically stream events from campus meters and sensors
($I_0,I_1$), have historical CSV meter data uploaded in bulk occasionally ($I_6$), and XML documents
fetched from NOAA weather services ($I_7$). Data from these sources are parsed by a pellet to
extract information ($I_7$), another pellet semantically annotates the event tuples to provide
context ($I_3$), and others insert/update these semantic triples to a 4Store semantic database
($I_4,I_8,I_9$). The ingest progress is available by an output pellet ($I_5$).

This pipeline exhibits many of the \floe design patterns, including streamed execution pull ($I_1$),
interleaved merge ($I_2$), switch control flow ($I_3$), and single execution push ($T_5$). Many
pellets use single messages but some use map tuples ($I_3$). Task parallelism is apparent, and some
($I_2,I_4$) allow multiple pellet instances for data parallel execution.  This \floe graph, in
active use on USC campus, runs on the \texttt{Tsangpo} private Eucalyptus Cloud having 128~cores in
16~nodes. The graph uses containers on 7~Extra Large VM instances (8~cores and 16~GB RAM each), and
by default, one flake is assigned to one container with dynamic adaptation strategy. Two additional
VMs host the \floe coordinator and manager. 





\subsection{Distributed Online Stream Clustering}
Stream clustering is an important precursor for many real-time microblog post and tweet analytics
algorithms. In Smart Grids, it can help utilities understand energy efficiency trends, and help
recommend appropriate customers for demand reduction. Stream clustering distributes posts from
incoming microblog feeds, which have a mix of topics, into output topic streams that groups posts
with a high likelihood of being in a topic. A distributed version of Locality Sensitive Hashing
(LSH)~\cite{gionis1999similarity} algorithm for stream clustering can be composed as a \floe graph. LSH is a family
of hash functions $H= \lbrace h : S \rightarrow  U\rbrace$ such that $\forall p,q: p \in B(q,r1) \Rightarrow
 Pr_H[h(q) = h(p)] \geq p_1$ and $\forall p,q: p \notin B(q,r2) \Rightarrow Pr_H[h(q) = h(p)] \leq
 p2$, for some $p1,p2,r1,r2$.  In other words, if two points are close together, then they remain
 close after a projection operation using a LSH, and these hash functions can significantly reduce
 the nearest neighbor search space.  

Fig.~\ref{fig:sampleapps}(b) shows the \floe graph for stream clustering operating on streams from
various news feeds. Text Cleaning pellet ($T_0$) cleans the posts using stemming, stop-word removal,
and spell-check to generate a feature vector based on dictionary of topic words. The Bucketizer
($T_1,T_2$) applies LSH to the vector to generate a several hash values that indicate stream buckets
with a high similarity to the post. We use \floe's dynamic data mapping pattern, that is similar to
but more versatile than MapReduce, to continuously route and automatically group posts to specific
Cluster Search pellets ($T_3,T_4,T_5$). Data parallel pellet instances of Cluster Search find the
closest locally matching cluster among the candidate buckets, acting as a local ``combiner'' on the
reducer's output. These locally closest results are sent to the Aggregator pellet ($T_6$) to find
the global best cluster for the post. A further feedback loop (with choice) notifies one of the
Cluster Search pellets of the updated post in its bucket, to be include in future comparisons. The
dynamic data mapping and loop offered by \floe for continuous processing makes this an elegant and
efficient solution for low-latency online analytics.





\subsection{Validation of Dynamic Allocation Strategies}

\begin{figure*}%
\centering
\subfigure[Message Count in Input Queue (Y1) \& Incoming Data Rate (Y2) for Pellet $I_1$ over Time]{
\includegraphics[width=\textwidth]{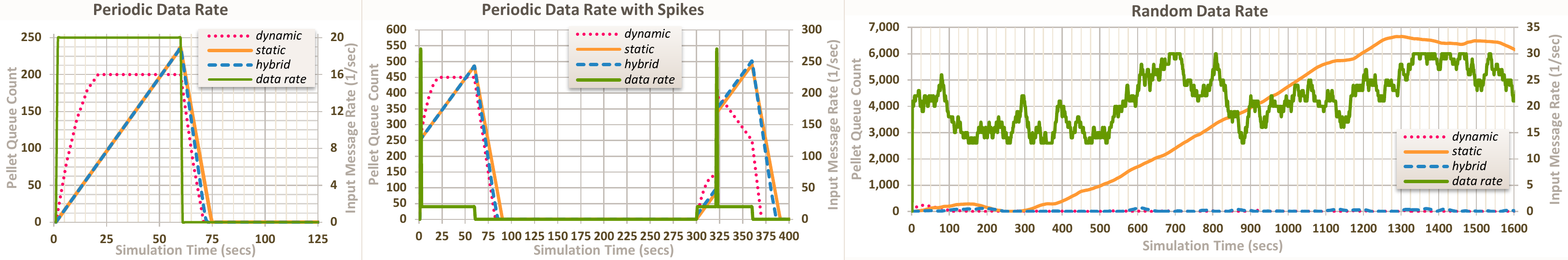}
\label{fig:eval-queue}
} 
\\
\subfigure[Number of Core Allocations for Pellet $I_1$ over Time]{
\includegraphics[width=\textwidth]{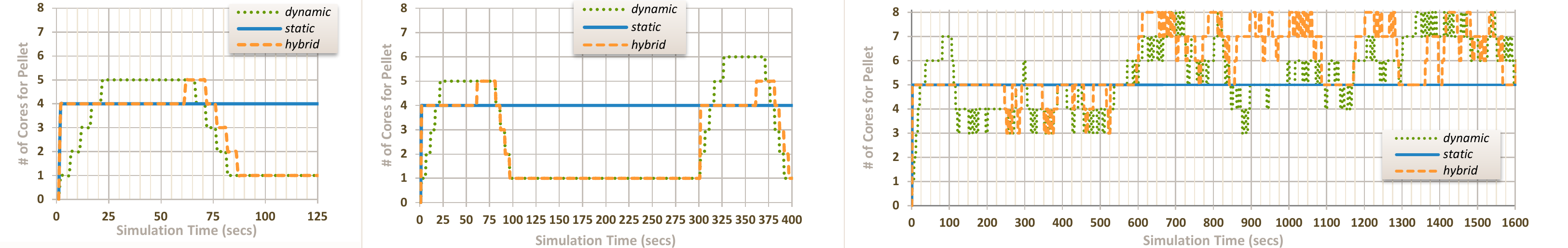}
\label{fig:eval-core}%
}
\caption{Simulation plots show resource behavior of pellet $I_1$ for periodic, periodic w/ spikes
  and random data rates.}
\vspace{-0.25in}
\label{fig:sim-eval}
\end{figure*}


The \floe framework currently supports static and dynamic adaptation, but not hybrid. To perform a
uniform comparison, simulate the Information Integration Pipeline (Fig.\ref{fig:sampleapps}(a)) and
compare the resource adaptation strategies under different load profiles of input messages entering
the dataflow at $I_0$. We focus on three profiles observed in our applications: \emph{periodic} with a
constant data rate, \emph{random} data rate with fluctuations similar to a one dimensional random
walk, and \emph{periodic with spikes} at random in the data rate. The $I_1$ pellet is used
representatively in the discussions, and we use a period of 5~mins and data duration of 60~secs for
the periodic data rates (i.e. 1~min data stream, 4~mins of gap).

Fig. \ref{fig:eval-core}(left) shows the number of allocated cores for over simulation time as it is
changed by the different strategies based on the \textbf{periodic} data rate. The static look-ahead
strategy allows a user-defined tolerance latency ($\epsilon$=20~secs) to complete data processing in
a period, on top of the data duration, and allocates the minimum number of cores to meet this
latency goal. Fig. \ref{fig:eval-queue}(left) shows this threshold of 80~secs being met at
75~secs, when the pellet's input queue has been drained. The dynamic strategy gradually allocates enough cores to achieve a steady state processing
(output data rate = input data rate) and hence finishes earlier at 70~secs, but at the cost of some
extra resources in that duration as seen by the area under the curve in
Fig. \ref{fig:eval-core}(left). For the periodic workload profile, the hybrid strategy performs
similar to the static look-ahead strategy, but additionally quiesces to 0 cores once done
processing, like the dynamic strategy.

For \textbf{periodic with random spikes} in the data rate, the static look-ahead strategy misses the
specified latency tolerance since it optimizes for the expected periodic data and is unable to
handle the unexpected data surges (Fig.~\ref{fig:eval-queue}(center)).  The dynamic strategy is able
to process all messages within the given tolerance while having a larger peak in core allocation
(Fig.~\ref{fig:eval-core}(center)). However, this is justified by its handling the data burst and
gracefully draining its input queue, as seen by the flat message count level. The hybrid model
performs better than the static model and finishes processing the messages within the given
tolerance while using less resources than the dynamic model. 

The last load profile has a \textbf{random} data rate with a known long-term average rate and slow
variation. (Fig.~\ref{fig:eval-queue}(right)). We observe that the static look-ahead strategy that
optimizes for only for the expected average data rate performs poorly and its input queue length
(and hence the queuing latency) accumulates over time. In contrast, the dynamic and hybrid strategy
adapt the number of cores dynamically to the data rate and are able to keep the pending input
messages negligible. In addition, the ratio of cumulative resources (area under the curves in
Fig.\ref{fig:eval-core}(right)) used by the static, the dynamic and the hybrid strategies is
$0.87:1.00:0.98$. This indicates that while the static look-ahead uses almost the
same resources as the other two, it performs much worse. 

Hence, we can surmise that the hybrid approach gives the broadest measure of good results, in terms
of limiting violation of the latency, while conserving the resources used by the pellets, under
different load profiles.


\section{Related Work}



Scientific workflow systems such as Kepler/ COMAD \cite{Ludscher2005} and Pegasus
\cite{Deelman2002}  have explored different
facets of distributed and data intensive processing of control and data flow applications. Most of these systems have been developed for one-shot batch processing workflows processing using files or objects while some such as Kepler/COMAD also support operations over collections. There has also been some work on incorporating stream processing into these workflow systems such as Kepler \cite{zinn:ccgrid:2010} and  Confluence \cite{Neophytou2011}, however these systems lack support for rich programming abstractions for stream based execution, especially support for dynamic key mapping (MapReduce) and BSP models that are supported by \floe.

Systems such as granules \cite{pallickara2009granules}, on the other hand, focus on particular abstractions such as MapReduce but fail to provide a generic stream processing model that can integrate various dataflow patterns with advanced patterns such as MapReduce. 

S4\cite{neumeyer2010s4} and IBM Stream processing core (SPC)\cite{amini2006spc}  provide a very similar distributed stream processing environment as \floe. It also provides generic programming model where the use can define the processing elements, similar to \floe's pellets, that can be composed to form a continuous dataflow application. However, unlike \floe, these system lack elastic scaling based on the dynamic data rate. 

There is yet another body of work on stream processing for continuous data streams and events, dating back to work on the TelegraphCQ\cite{chandrasekaran2003telegraphcq}  and Aurora \cite{aurora}. More recently, Schneider et. al. \cite {Schneider2009} developed a system built on top of the SPADE system to allow Operator Elasticity, wherein the standard operators perform their computations in parallel, while also dynamically adapting parallelism level to obtain the best performance based on the current system load. These optimizations are similar to the dynamic resource allocation strategy in \floe, however, in addition we exploit the estimated data characteristics though the static look-ahead and hybrid strategies that minimizes resource utilization while conforming to the latency requirements.

StreamCloud\cite{Gulisano2010} is another stream processing system that focuses on scalability with respect to the stream data rates. StreamCloud achieves this by partitioning the incoming data stream with semantic awareness about the operator in the downstream instance and achieves scalability with intra-operator parallelism and dynamic resource allocation. Esc \cite{Satzger2011} is a novel elastic stream processing platform that offers a streaming model using Cloud resources on-demand to adapt to changing computational demands.  Esc allows the end user to specify adaptation strategies in terms
of data partitioning as well splitting individual tasks into smaller parallel tasks based on the workload. The optimizations in \floe do not assume any pellet semantics and the optimizations are performed implicitly by the framework by exploiting the data parallel nature of the operations or explicitly by the application composer by using the data based choice or round robin split patterns for load balancing.

Application dynamism have be explored using concepts such as frames, templates and dynamic embedding\cite{Ngu2008}. However, these are in the context of scientific workflows where the choice for the dynamic task is finalized before the execution of that task begins which is in contrast to the requirement in the continuous dataflows where an executing task needs to be seamlessly updated at runtime as is support by \floe.

\section{Conclusions}

In this article, we have motivated the needs of an emerging class of dynamic, continuous dataflow
applications, that are unique in the application and dynamism they exhibit, and the rich composition
patterns they use. Our work has explored the design paradigms for a Cloud-based framework, \floe, to
meet these needs, with elastic optimizations to meet variable streaming data rates. Our future work
will pursue research novel problems related to resiliency of these applications, and additional
optimizations evaluated under real Cloud environments. 


\bibliographystyle{IEEEtran}
\bibliography{main}

\end{document}